\documentclass[conference]{IEEEtran}
\IEEEoverridecommandlockouts
\usepackage{cite}
\usepackage[breaklinks=true]{hyperref}
\usepackage{breakcites}
\usepackage{amsmath,amssymb,amsfonts}
\usepackage{algorithmic}
\usepackage{graphicx}
\usepackage{textcomp}
\usepackage{xcolor}
\usepackage{subfig}
\usepackage{algorithm}
\usepackage{algorithmic}
\usepackage{amsmath}
\usepackage{lipsum}
\usepackage{array}
\usepackage{xcolor}
\usepackage{booktabs}

\usepackage{listings}
\def\BibTeX{{\rm B\kern-.05em{\sc i\kern-.025em b}\kern-.08em
 T\kern-.1667em\lower.7ex\hbox{E}\kern-.125emX}}

\begin{document}

\title{A monitoring system for collecting and aggregating metrics from distributed clouds \\
\thanks{\includegraphics[width=0.4cm]{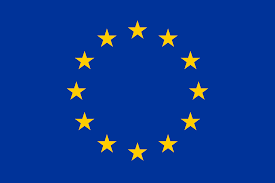} Funded by the European Union (TaRDIS, 101093006). Views and opinions expressed are however those of the author(s) only and do not necessarily reflect those of the European Union. Neither the European Union nor the granting authority can be held responsible for them.}
}

\makeatletter 
\newcommand{\linebreakand}{%
  \end{@IEEEauthorhalign}
  \hfill\mbox{}\par
  \mbox{}\hfill\begin{@IEEEauthorhalign}
}
\makeatother 

\author{\IEEEauthorblockN{Tamara Ranković}
\IEEEauthorblockA{\textit{Faculty of Technical Sciences} \\
University of Novi Sad \\
Novi Sad, Serbia \\
tamara.rankovic@uns.ac.rs}
\and
\IEEEauthorblockN{Mateja Rilak}
\IEEEauthorblockA{\textit{Faculty of Technical Sciences} \\
University of Novi Sad \\
Novi Sad, Serbia \\
rilak.sr38.2021@uns.ac.rs}
\and
\IEEEauthorblockN{Janko Rakonjac}
\IEEEauthorblockA{\textit{Faculty of Technical Sciences} \\
University of Novi Sad \\
Novi Sad, Serbia \\
rakonjac.sr33.2021@uns.ac.rs}
\and
\IEEEauthorblockN{Miloš Simić}
\IEEEauthorblockA{\textit{Faculty of Technical Sciences} \\
University of Novi Sad \\
Novi Sad, Serbia \\
milos.simic@uns.ac.rs}
}

\maketitle

\begin{abstract}
Applications requiring real-time processing of large volumes of data have been the main driver for rethinking the traditional cloud, giving rise to novel cloud models.
Distributed cloud (DC) is a model that allows users to dynamically create and dispose of strategically located ad-hoc clouds that contain resources best tailored to their needs.
It is essential for this model to provide a high degree of observability for it to be viable in real-world scenarios.
In this paper, we present the design and implementation of a monitoring system that collects metrics from DCs and makes them accessible to diverse clients.
Agents running on nodes are responsible for collecting machine-, container-, and application-level metrics.
During the health-check protocol, that data is transferred from the node to DC's control plane running inside the cloud.
There, it is persisted and served via multiple APIs, including a streaming API.
Moreover, node metrics are aggregated for every DC in order to provide a more comprehensive view of the system's state.
\end{abstract}


\section{Introduction}

In the past decade, we have seen a pronounced transformation of the cloud model.
Requirements posed by modern applications have been the main driver of this transformation.
Real-time transfer, processing, and retrieval of large volumes of data have been shown to be inefficient or impossible when relying on remote centralized datacenters \cite{ferrer2019towards}.
To address these issues, many novel cloud models, like edge and fog computing, have emerged \cite{shi2016edge, bonomi2012fog}.
While proposing different solutions, they all embrace the concept of data locality - processing data on locations where it is stored, but also storing it near locations where it is generated.
Modern manufacturing facilities, power plants, autonomous vehicles, and other related use cases, where data sources and consumers are in near proximity and real-time data processing and retrieval are essential, benefit from this organization tremendously.
Besides introducing efficiency, data privacy and security are also enhanced, as data may never cross the boundary of trusted locations or networks.

The distributed cloud (DC) model is one of the models concerned with presented problems \cite{simic2021towards, simic2021infrastructure}.
It consists of cloud and DC layers.
The cloud layer is the traditional centralized cloud, while the DC layer is meant to offload the cloud layer and also serve clients with strict latency and privacy requirements.
The DC layer consists of dynamic DCs formed and discarded by users.
Each DC should be tailored to the user's needs, including its resource capacity and location \cite{simic2021towards, greenberg2008cost}.
The lifecycle of a DC is coordinated by a control plane in the cloud layer, but the DC itself ensures its proper operation, providing a hybrid architecture.
Users are expected to specify configuration and workloads for their DCs, while the system is responsible for its convergence to the desired state.
This model allows for seamless cooperation between the cloud and the DC.

To make the DC model a viable option for real-world applications, we need to provide a rich ecosystem of functionalities and reliable tools around it.
Efforts towards this have been made by introducing namespaces and dataspaces, allowing for virtual DCs that provide multitenancy capabilities \cite{simic2024hierarchical, simic2025data}.
Moreover, mechanisms for dynamic system and application reconfiguration were developed \cite{rankovic2024configuration}, with native support for misconfiguration prevention and detection \cite{rankovic2024misconfiguration}.
Access to any resource in the DC is managed by a hierarchical attribute-based model \cite{rankovic2024access, rankovic2023enforcing}.

Considering DC's complex architecture and interactions among its components, it is necessary to ensure a high degree of observability.
To achieve this in any environment, it is a standard practice to collect various metrics, logs, and traces \cite{karumuri2021towards}.
They serve as a foundation for monitoring the system's behavior and performance and detecting any bottlenecks or failures.
Moreover, workload schedulers and infrastructure autoscalers rely primarily on metrics as inputs for making decisions.

Existing monitoring solutions have become highly sophisticated and have numerous capabilities, however, none of them are native to the DC model.
As DCs are dynamic, regularly being formed or destroyed, and with many nodes constantly joining or leaving, the monitoring system needs to be flexible enough to accommodate these fluctuations.
Additionally, as hardware comprising DCs can often be resource-limited, lightweight tools are essential to ensure minimal resource consumption by system components.

The focus of this paper is on designing and implementing a monitoring system native for DCs, more precisely for collecting metrics from DCs and making them accessible to clients.
We consider end users who manage DCs the primary clients, as it is crucial for them to have insight into the state of their DC infrastructure and applications running on it.
However, we take more potential clients into consideration, including workload schedulers, infrastructure, and application-level autoscalers, who make decisions based on this data.
Moreover, if these clients incorporate machine learning models, they can utilize collected metrics also as training sets.

In Section \ref{s:rw} we discuss existing monitoring solutions, particularly the cloud-native ones.
Sections \ref{s:sd} and \ref{s:i} present our solution, its design and implementation respectively.
In Section \ref{s:d} we provide reasoning behind certain design decisions and discuss the system's limitations and possible improvements, while Section \ref{s:c} concludes the paper and outlines directions for future work.

\section{Related work}
\label{s:rw}

Prometheus\footnote{Prometheus: \url{https://prometheus.io/}} is a widely used monitoring system with a rich set of tools and integrations.
It is a centralized metrics collection and analysis solution that uses a pull communication model.
A Prometheus instance is configured to periodically scrape metrics from desired targets and store them in its time-series database.
Targets can expose their metrics directly when instrumented with client libraries.
However, when that is not possible, usually in third-party systems, numerous exporters have been developed to extract and transform metrics from such systems into the Prometheus format.
The standardization of this format, called OpenMetrics\footnote{OpenMetrics: \url{https://openmetrics.io/}}, was introduced to encourage interoperability between different monitoring tools.

Kubernetes\footnote{Kubernetes: \url{https://kubernetes.io/}}, a popular container orchestrator, has a resource metrics pipeline and a full metrics pipeline.
The first one serves a limited set of metrics via the metrics-server which collects metrics from nodes and stores them only in memory for a short time.
This pipeline is primarily meant to be used by autoscalers.
The second pipeline provides a significantly richer set of metrics from nodes but requires users to implement their adapter that exposes those metrics.
All metrics are transmitted in the OpenMetrics format.

When orchestrating workloads over a federation of clusters, a scheduler needs to have a global view of the entire federation.
Acala \cite{huang2024aggregate} is designed to aggregate metrics from federations of Kubernetes clusters.
Global View Cluster contains Prometheus storage and a set of Acala controllers that periodically request metrics from clusters.
There is an Acala member in every cluster that processes these requests by collecting metrics from all nodes inside that cluster.
Before returning a response, it applies data reduction strategies, aggregation, and deduplication.
Aggregation calculates cluster averages for metrics that allow so, and discards values for individual nodes, while deduplication compares the current with the previous value of the same metric, and discards it if they are equal.

There have been efforts to integrate monitoring into cloud-extending models similar to the DC, so we give them special attention as solutions developed for one model could also apply to the others \cite{costa2022monitoring}.
EdgeCloud Mon \cite{korontanis2024edgecloud} integrates cloud with edge by proposing a system that collects infrastructure, pod, and virtual machine metrics from k3s\footnote{k3s: \url{https://k3s.io/}} edge clusters and transfers them to the cloud for long-term storage and access.
Every node exposes infrastructure metrics through the node exporter\footnote{Node exporter: \url{https://github.com/prometheus/node_exporter}} and a custom agent that provides additional machine metrics.
Additionally, kubevirt-exporter collects metrics for virtual machines running on worker nodes by collecting them via the KubeVirt API.
This virtualization technique is aimed at workloads that require Windows as the operating system.
Inside every cluster, a Prometheus instance periodically scrapes metrics from exporters on worker nodes and kubevirt-exporter running on the master node, and pushes them to the cloud.

Compared to all the centralized solutions we discussed, FogMon \cite{brogi2019measuring} presents a two-layer distributed architecture that promotes fault tolerance.
On the first layer are follower nodes that collect their metrics and send them to the leader node they were assigned to, which resides in the second layer.
Leader nodes aggregate metrics from follower nodes and share them with other leaders through a peer-to-peer network via a gossip protocol.

\begin{figure*}[t!]
    \center
    \includegraphics[width=0.9\textwidth]{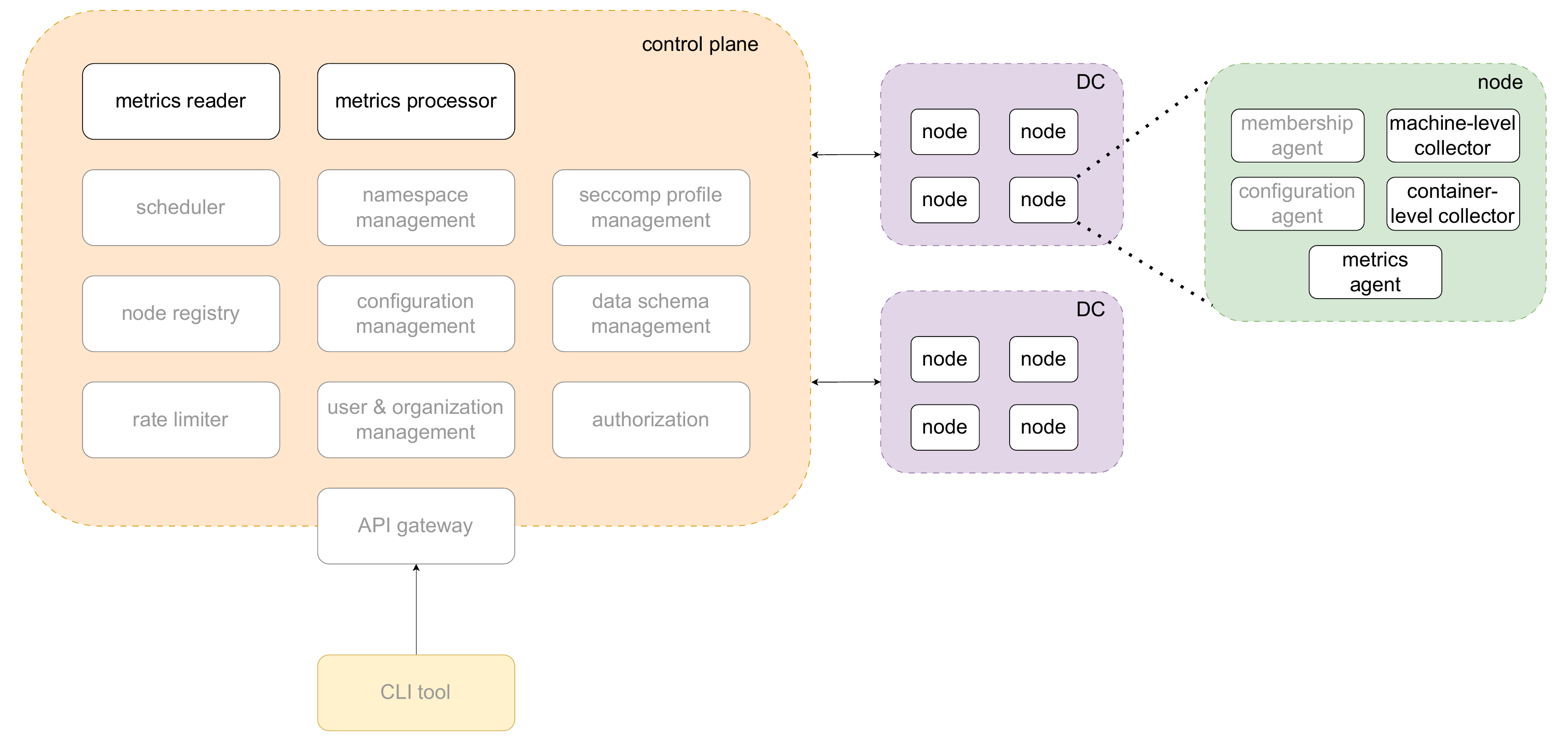}
    \caption{DC platform architecture}
    \label{fig:a}
\end{figure*}

\section{System design}
\label{s:sd}

To provide a comprehensive overview of a DC's state, we need to collect metrics on multiple levels.
The first one is the \emph{machine level}, consisting of each node's hardware and system software.
Aside from gathering machine metrics for individual nodes, it is also valuable to aggregate a subset of them for every DC.
Therefore, we differentiate between node-level and DC-level machine metrics.
The second level is the \emph{container level}, as all user-defined workloads are running inside this isolated environment.
When the model evolves to support other environments, such as virtual machines and unikernels, this level will also evolve to incorporate them, so we may refer to it as the \emph{isolated environment level}.
Lastly, the \emph{application level} is the one that should allow client applications to generate arbitrary metrics that will be processed by and stored in the system.

The metrics subsystem comprises multiple components responsible for collecting, processing, storing, and retrieving metrics.
They can be separated into ones running on nodes and ones running in the control plane.
Fig. \ref{fig:a} displays the extended architecture of the DC platform.

Every node contains the following components:
\begin{itemize}
    \item \textbf{Machine-level collector} providing an API for accessing node-level machine metrics.
    \item \textbf{Container-level collector} providing an API for accessing container-level metrics.
    \item \textbf{Metrics agent} responsible for collecting node metrics from all levels and transferring them to the control plane.
\end{itemize}

Inside the control plane, we have another set of components:
\begin{itemize}
    \item \textbf{Metrics storage} where all collected metrics reside.
    \item \textbf{Processor} responsible for collecting metrics from the nodes, aggregating and storing them.
    \item \textbf{Reader} implementing the API for retrieving the collected metrics, intended for clients of the DC platform.
\end{itemize}

The collection process can be divided into two steps, the first one being the collection of metrics from different levels inside a node, and the second one being the collection from nodes by the control plane.
Firstly, we discuss how metrics collection on the node is designed.

Machine and container metrics are exposed by their respective collectors that implement an agreed-upon API so that their implementation can be swapped seamlessly.
The API must have only one HTTP GET /metrics endpoint, following the example of Prometheus.
A response to the endpoint invocation should be a list of collected metrics from the previous invocation.
Here, and in all other metrics transmission points, we comply with the OpenMetrics standard, in order to simplify integration with existing cloud-native tools.
The metrics agent periodically sends requests to machine and container collectors and stores results on the file system temporarily, until they are transferred to the control plane for long-term persistence.
The frequency at which the agent performs this operation is configurable.

Application-level metrics operate similarly, with the main difference being that collection targets are dynamic, and defined by DC users.
Therefore, the Metrics agent must maintain a target list.
When a user submits a new application to the control plane, they should specify if that application implements a metrics API, the same one described for machine and container collectors.
If it does, when the scheduler assigns an instance of that application to a node, it should notify the Metrics agent of the address where the application will be reachable.
When the application gets terminated, the agent should be notified once again, so it can remove the address from its target list.
As with metrics on other levels, the agent periodically polls the targets, gathers results, and stores them on a file system.
It should be noted that the system doesn't restrict what metrics are to be collected.
They are completely custom, allowing users to generate any information they deem valuable from their applications, as long as the API is implemented correctly.

Now we describe how metrics are transfered from the nodes to the control plane and are processed and stored there.
Processor is a component whose main purpose is to monitor the health of the node pool.
It periodically sends a ping request and, if a node doesn't reply within a defined interval, it is considered dead.
The processor relies on the node registry to get the list of all nodes and updates it according to health-check results.
We piggyback metrics data on this protocol, with the goal of reducing the number of requests between nodes and the control plane.
When a node replies to a ping, it also adds to the reply all metrics collected from the previous ping.
After that, it clears transmitted data from its temporary storage.
When the Processor receives metrics, it writes them to the Metrics storage.
Even though this method can cause data loss if the Processor doesn't successfully persist the data, we opted for it because one of the main properties of nodes is that they are resource-limited.

As we previously mentioned, certain machine metrics, such as total and available hardware resources, should exist not only on a node level but also on a DC level.
A simple strategy we propose is for a Processor to periodically retrieve the latest node-level metrics for each DC, perform a respective aggregation function, usually a sum or an average, and store the result.

The Reader component implements APIs for clients to access stored metrics.
It provides both a REST and a streaming API.
The REST API endpoints are listed in Table \ref{t:1}.
In all cases, the response is a list of metrics in the OpenMetrics text format.

\setlength{\tabcolsep}{0.5em}
{\renewcommand{\arraystretch}{2}
\begin{table*}[h]
    \caption{Metrics REST API}
    \label{t:1}
    \begin{center}
    \begin{tabular}{m{1cm}m{5cm}m{4cm}m{6cm}p{0pt}}
    \toprule
    \center \textbf{Method} & \center \textbf{Path}& \center \textbf{Parameters}& \center \textbf{Description} &\\
    \midrule
    &\vfill /api/metrics-api/nodes/\{nodeID\}/\{timestamp\}&\vfill nodeID - Node identifier\newline timestamp - Range start timestamp&The endpoint retrieves metrics for the specified node, collected from the specified timestamp until now. \\
    \cline{2-4} 
    &\vfill /api/metrics-api/nodes/\{nodeID\}/\{start\}/\{end\}&\vfill nodeID - Node identifier\newline start - Range start timestamp\newline stop - Range stop timestamp&The endpoint retrieves metrics for the specified node, collected for the specified time range. \\
    \cline{2-4}
    &\vfill /api/metrics-api/dc/\{dcID\}/\{timestamp\}&\vfill dcID - DC identifier\newline timestamp - Range start timestamp&The endpoint retrieves metrics for the specified DC, collected from the specified timestamp until now. \\
    \cline{2-4} 
    &\vfill /api/metrics-api/dc/\{dcID\}/\{start\}/\{end\}&\vfill dcID - DC identifier\newline start - Range start timestamp\newline stop - Range stop timestamp&The endpoint retrieves metrics for the specified DC, collected for the specified time range. \\
    \cline{2-4} 
    &\vfill /api/metrics-api/containers/\{containerID\}/\{timestamp\}&\vfill containerID - Container identifier\newline timestamp - Range start timestamp&The endpoint retrieves metrics for the specified container, collected from the specified timestamp until now. \\
    \cline{2-4} 
    &\vfill /api/metrics-api/containers/\{containerID\}/\{start\}/\{end\}&\vfill containerID - Container identifier\newline start - Range start timestamp\newline stop - Range stop timestamp&The endpoint retrieves metrics for the specified container, collected for the specified time range. \\
    \cline{2-4} 
    \center GET&\vfill /api/metrics-api/apps/\{appID\}/\{timestamp\}&\vfill appID - Application identifier\newline timestamp - Range start timestamp&The endpoint retrieves metrics for the specified application, collected from the specified timestamp until now. \\
    \cline{2-4} 
    &\vfill /api/metrics-api/apps/\{appID\}/\{start\}/\{end\}&\vfill appID - Application identifier\newline start - Range start timestamp\newline stop - Range stop timestamp&The endpoint retrieves metrics for the specified application, collected for the specified time range. \\
    \cline{2-4}
    &\vfill /api/metrics-api/\{timestamp\}&\vfill timestamp - Range start timestamp&The endpoint retrieves all metrics collected from the specified timestamp until now. Primarily intended to be used for training of machine learning models. \\
    \cline{2-4} 
    &\vfill /api/metrics-api/\{start\}/\{end\}&\vfill start - Range start timestamp\newline stop - Range stop timestamp&The endpoint retrieves all metrics collected for the specified time range. Also intended to be used for training machine learning models. \\
    \cline{2-4} 
    &\vfill /api/metrics-api/nodes/\{nodeID\}/latest&\vfill nodeID - Node identifier&The endpoint retrieves latest metrics for the specified node. It contains only a subset of crucial metrics, intended for displaying on dashboards. \\
    \cline{2-4} 
    &\vfill /api/metrics-api/dc/\{dcID\}/latest&\vfill dcID - DC identifier&The endpoint retrieves latest metrics for the specified DC. It contains only a subset of crucial metrics, intended for displaying on dashboards. \\
    \bottomrule
    \end{tabular}
    \label{tab1}
    \end{center}
\end{table*}

The streaming API allows clients to subscribe to the latest node or DC metrics.
It can be useful for real-time dashboards, schedulers, or autoscalers.
We define a topic hierarchy \emph{metrics.\{nodes\textbar dc\}.\{id\textbar *\}} and allow the use of wildcards (*) to subscribe to multiple topics at once.
If a client wants to subscribe to metrics of a DC with id X, they should use the topic \emph{metrics.dc.X}.
On the other hand, if they want to receive metrics for all nodes, the topic would be \emph{metrics.nodes.*}

\section{Implementation}
\label{s:i}

The DC platform we provided with a monitoring subsystem is an open-source project named \emph{constellations (c12s)}\footnote{c12s source code: \url{https://github.com/c12s/}}.
Starometry\footnote{starometry source code: \url{https://github.com/c12s/starometry}} is an implementation of the Metrics agent, while Processor and Reader are both part of the protostar project\footnote{protostar source code: \url{https://github.com/c12s/protostar}}.

All components, including the ones presented in this paper, are written in Go programming language.
Communication between the control plane and nodes, as well as the streaming API, are implemented using NATS\footnote{NATS: \url{https://nats.io/}} as the message broker.
Node exporter and Windows exporter\footnote{Windows exporter: \url{https://github.com/prometheus-community/windows_exporter}} are used as machine-level collectors on nodes, while cAdvisor\footnote{cAdvisor: \url{https://github.com/google/cadvisor}} is used for accessing container-level metrics.
Prometheus was chosen as the Metrics storage, because of its native support for OpenMetrics format and a rich query language.

The correctness of the implementation was tested in laboratory conditions with simulated nodes running inside virtual machines.
However, tests in realistic environments on heterogeneous hardware are part of our future work.

\section{Discussion}
\label{s:d}

In this section, we will present the reasoning behind the system's design choices and discuss our solution's limitations, as well as directions for possible improvements.

When deciding on the architecture, we primarily focused on small-scale DCs, where centralized collection and processing points are sufficient.
However, in large DCs comprising thousands of nodes, this strategy can easily overload the control plane and cause more frequent failures or traffic congestion.
For these scenarios, we propose a peer-to-peer network to be formed inside every DC.
With communication established between nodes, they would be able to perform health checks internally, and the control plane needs to directly communicate with only a small number of nodes to collect states for the entire DC.
Furthermore, metrics can still be piggybacked on the health-check protocol, so that data aggregation load can be distributed between nodes.
There are numerous approaches to this, from gossip-based aggregation \cite{kempe2003gossip, jelasity2005gossip}, to hierarchical strategies \cite{massie2004ganglia, graffi2017skyeye}.
We leave this question to our future work.

As nodes generate large amounts of metrics data constantly, it is necessary to incorporate strategies to reduce it in transit and at rest.
Here we distinguish between lossless and lossy strategies.
The usage of lossless compression algorithms suitable for time-series data could significantly lower the amount of network and persistent storage required.
However, the focus should be on lightweight algorithms, as compression and decompression would introduce latency and the need for more processing power.
A lossy strategy that could drastically decrease data size is a dynamic sampling of metrics so that only relevant data points are recorded.
If values are stable over long periods, with occasional notable changes, most of that data can be discarded on the node.
This way, only a small subset has to be transferred to the control plane and permanently stored, while the rest can be reconstructed with a satisfactory degree of accuracy. 

One design decision we mentioned in Section \ref{s:sd}, that could present a system limitation in certain cases, is potential data loss if a node sends its metrics, and deletes it, but the control plane doesn't persist it successfully.
We opted for this approach to reduce the node's disk usage and with an assumption that rare metrics losses are tolerable to users.
However, if nodes have ample disk space and no metrics loss is allowed, the proposed collection strategy can be altered.
The control plane could be extended to send confirmation messages when it saves the data, and only then would nodes remove it from its storage.
If no confirmation is received within a defined interval, the node will resend that data.

Because DCs should support diverse use cases and can have varying structures, they have to offer multiple strategies regarding metrics collection, aggregation, compression, and storage.
Therefore, we argue that the platform should include all the approaches we discussed here and allow the end user to select a configuration most appropriate for their needs.

\section{Conclusion}
\label{s:c}

In this paper, we proposed one possible design for a metrics collection, aggregation, and retrieval system operating inside DCs, and discussed its refinements and alternatives.
It covers metrics from the lowest machine level, to the custom metrics generated by client applications.
Components are separated into ones running on nodes, responsible for data collection, and ones inside the control plane, gathering data from nodes, aggregating, and serving it to users.
We supplemented this design with an implementation integrated into an existing open-source DC platform.

As our solution is centralized, heavily relying on the control plane, we plan to complement it with a more decentralized strategy in which DCs collect and aggregate data, and only later share it with the control plane for easier access.
Also, we will focus on lightweight algorithms that filter and compress metrics in order to minimize the amount of required resources.
Furthermore, a well-equipped alerting mechanism is an indispensable component of any monitoring system, thus its development is also part of our future work.
We also plan to utilize the monitoring system for developing infrastructure and application-level autoscaleres.
To ensure the correct and efficient operation of all components, we will test our system in realistic environments.

\bibliographystyle{bib/IEEEtran}
\bibliography{bib/refs}

@article{ferrer2019towards,
  title={Towards the decentralised cloud: Survey on approaches and challenges for mobile, ad hoc, and edge computing},
  author={Ferrer, Ana Juan and Marqu{\`e}s, Joan Manuel and Jorba, Josep},
  journal={ACM Computing Surveys (CSUR)},
  volume={51},
  number={6},
  pages={1--36},
  year={2019},
  publisher={ACM New York, NY, USA}
}

@article{shi2016edge,
  title={Edge computing: Vision and challenges},
  author={Shi, Weisong and Cao, Jie and Zhang, Quan and Li, Youhuizi and Xu, Lanyu},
  journal={IEEE internet of things journal},
  volume={3},
  number={5},
  pages={637--646},
  year={2016},
  publisher={Ieee}
}

@inproceedings{bonomi2012fog,
  title={Fog computing and its role in the internet of things},
  author={Bonomi, Flavio and Milito, Rodolfo and Zhu, Jiang and Addepalli, Sateesh},
  booktitle={Proceedings of the first edition of the MCC workshop on Mobile cloud computing},
  pages={13--16},
  year={2012}
}

@article{simic2021towards,
  title={Towards edge computing as a service: Dynamic formation of the micro data-centers},
  author={Simi{\'c}, Milo{\v{s}} and Proki{\'c}, Ivan and Dedei{\'c}, Jovana and Sladi{\'c}, Goran and Milosavljevi{\'c}, Branko},
  journal={IEEE Access},
  volume={9},
  pages={114468--114484},
  year={2021},
  publisher={IEEE}
}

@article{simic2021infrastructure,
  title={Infrastructure as software in micro clouds at the edge},
  author={Simi{\'c}, Milo{\v{s}} and Sladi{\'c}, Goran and Zari{\'c}, Miroslav and Markoski, Branko},
  journal={Sensors},
  volume={21},
  number={21},
  pages={7001},
  year={2021},
  publisher={MDPI}
}

@article{simic2024hierarchical,
  title={A Hierarchical Namespace Approach for Multi-Tenancy in Distributed Clouds},
  author={Simi{\'c}, Milo{\v{s}} and Dedei{\'c}, Jovana and Stojkov, Milan and Proki{\'c}, Ivan},
  journal={IEEE Access},
  year={2024},
  publisher={IEEE}
}

@article{simic2025data,
  title={Data overlay mesh in distributed clouds allowing collaborative applications},
  author={Simi{\'c}, Milo{\v{s}} and Dedei{\'c}, Jovana and Stojkov, Milan and Proki{\'c}, Ivan},
  journal={IEEE Access},
  year={2025},
  publisher={IEEE}
}

@inproceedings{rankovic2024access,
  title={Access Control in a Distributed Micro-cloud Environment},
  author={Rankovi{\'c}, Tamara and Simi{\'c}, Milo{\v{s}} and Stojkov, Milan and Sladi{\'c}, Goran},
  booktitle={Conference on Information Technology and its Applications},
  pages={435--447},
  year={2024},
  organization={Springer}
}

@inproceedings{rankovic2024configuration,
  title={Configuration management in the distributed cloud},
  author={Rankovi{\'c}, Tamara and Kova{\v{c}}evi{\'c}, Ivana and Maksimovi{\'c}, Veljko and Sladi{\'c}, Goran and Simi{\'c}, Milo{\v{s}}},
  booktitle={Conference on Information Technology and its Applications},
  pages={224--235},
  year={2024},
  organization={Springer}
}

@inproceedings{rankovic2024misconfiguration,
  title={Misconfiguration prevention and error cause detection for distributed-cloud applications},
  author={Rankovi{\'c}, Tamara and {\v{S}}ilji{\'c}, Filip and Tomi{\'c}, Jovan and Sladi{\'c}, Goran and Simi{\'c}, Milo{\v{s}}},
  booktitle={2024 IEEE 22nd Jubilee International Symposium on Intelligent Systems and Informatics (SISY)},
  pages={000297--000302},
  year={2024},
  organization={IEEE}
}

@article{rankovic2023enforcing,
  title={Enforcing Zero Trust in Distributed Cloud Deployments},
  author={Rankovi{\'c}, T and Maksimovi{\'c}, V and Simi{\'c}, M and Milosavljevi{\'c}, B and Sladi{\'c}, G},
  journal={INFORMATION TECHNOLOGY AND DEVELOPMENT OF EDUCATION ITRO 2023},
  pages={142}
}

@article{huang2024aggregate,
  title={Aggregate Monitoring for Geo-Distributed Kubernetes Cluster Federations},
  author={Huang, Chih-Kai and Pierre, Guillaume},
  journal={IEEE Transactions on Cloud Computing},
  year={2024},
  publisher={IEEE}
}

@article{korontanis2024edgecloud,
  title={EdgeCloud Mon: A lightweight monitoring stack for K3s clusters},
  author={Korontanis, Ioannis and Makris, Antonios and Tserpes, Konstantinos},
  journal={SoftwareX},
  volume={26},
  pages={101675},
  year={2024},
  publisher={Elsevier}
}

@inproceedings{brogi2019measuring,
  title={Measuring the fog, gently},
  author={Brogi, Antonio and Forti, Stefano and Gaglianese, Marco},
  booktitle={International Conference on Service-Oriented Computing},
  pages={523--538},
  year={2019},
  organization={Springer}
}

@article{karumuri2021towards,
  title={Towards observability data management at scale},
  author={Karumuri, Suman and Solleza, Franco and Zdonik, Stan and Tatbul, Nesime},
  journal={ACM Sigmod Record},
  volume={49},
  number={4},
  pages={18--23},
  year={2021},
  publisher={ACM New York, NY, USA}
}

@article{costa2022monitoring,
  title={Monitoring fog computing: A review, taxonomy and open challenges},
  author={Costa, Breno and Bachiega Jr, Joao and Carvalho, Leonardo Rebou{\c{c}}as and Rosa, Michel and Araujo, Aleteia},
  journal={Computer Networks},
  volume={215},
  pages={109189},
  year={2022},
  publisher={Elsevier}
}

@inproceedings{kempe2003gossip,
  title={Gossip-based computation of aggregate information},
  author={Kempe, David and Dobra, Alin and Gehrke, Johannes},
  booktitle={44th Annual IEEE Symposium on Foundations of Computer Science, 2003. Proceedings.},
  pages={482--491},
  year={2003},
  organization={IEEE}
}

@article{jelasity2005gossip,
  title={Gossip-based aggregation in large dynamic networks},
  author={Jelasity, M{\'a}rk and Montresor, Alberto and Babaoglu, Ozalp},
  journal={ACM Transactions on Computer Systems (TOCS)},
  volume={23},
  number={3},
  pages={219--252},
  year={2005},
  publisher={ACM New York, NY, USA}
}

@article{graffi2017skyeye,
  title={SkyEye: A tree-based peer-to-peer monitoring approach},
  author={Graffi, Kalman and Disterh{\"o}ft, Andreas},
  journal={Pervasive and Mobile Computing},
  volume={40},
  pages={593--610},
  year={2017},
  publisher={Elsevier}
}

@article{massie2004ganglia,
  title={The ganglia distributed monitoring system: design, implementation, and experience},
  author={Massie, Matthew L and Chun, Brent N and Culler, David E},
  journal={Parallel Computing},
  volume={30},
  number={7},
  pages={817--840},
  year={2004},
  publisher={Elsevier}
}

@misc{greenberg2008cost,
  title={The cost of a cloud: research problems in data center networks},
  author={Greenberg, Albert and Hamilton, James and Maltz, David A and Patel, Parveen},
  journal={ACM SIGCOMM computer communication review},
  volume={39},
  number={1},
  pages={68--73},
  year={2008},
  publisher={ACM New York, NY, USA}
}

\end{document}